\documentclass[aps,prl,twocolumn,showpacs,nofootinbib]{revtex4-1}

\usepackage{bm,dcolumn,amsmath,graphicx}

\newcommand{\etal}{\textit{et al.}}

\begin{document}
\title{Comment on ``Global Positioning System Test of the Local Position Invariance of Planck's Constant''}

\author{J. C. Berengut}
\author{V. V. Flambaum}
\affiliation{School of Physics, University of New South Wales, Sydney, NSW 2052, Australia}

\date{26 March 2012}
\maketitle

In their Letter~\cite{kentosh12prl}, Kentosh and Mohageg seek to use data from clocks aboard global positioning system (GPS) satellites to place limits on local position invariance (LPI) violations of Planck's constant, $h$. It is the purpose of this comment to show that discussing limits on variation of dimensional constants (such as $h$) is not meaningful; that even within a correct framework it is not possible to extract limits on variation of fundamental constants from a single type of clock aboard GPS satellites; and to correct an important misconception in the authors' interpretation of previous Earth-based LPI experiments.

\paragraph{1.---}
It is not meaningful to discuss limits on $h$, $c$, or any other physical constant that contains dimension unless it is made clear what units are arbitrarily being held fixed. This is clear from the fact that it is always possible to choose units such that some of these constants are unity: e.g. $\hbar=c=1$ is a common choice among particle physicists. In SI units, $c$ is an integer number of metres per second and cannot vary since the metre is defined by $c$. Such freedom is not available for dimensionless constants. In any system of units one employs, the fine-structure constant $\alpha=e^2/\hbar c$ has a value of approximately $1/137$ (on Earth). One may therefore ask an LPI question: does the value of $\alpha$ at some point depend on the gravitational potential at that point? We parametrize the coupling of $\alpha$ to the gravitational potential $U$ by $\kappa_\alpha$:
\begin{equation}
	\alpha_x/\alpha_0 = 1 + \kappa_\alpha \Delta U/c^2 \,,
\end{equation}
where $\alpha$ is measured at points $x$ and $0$ separately~\cite{flambaum08aipconf,ferrell07pra,blatt08prl}. On the other hand Eq.~(6) from~\cite{kentosh12prl}, $h_x/h_0 = 1 + \beta_h \Delta U/c^2$, requires specification of units; in units of $e^2/c$ it is equivalent to (1).

\paragraph{2.---}
Tests of LPI violation can be performed using two different clocks on the same satellite. To remove the effect of general relativity one needs to measure the dimensionless ratio of their frequencies, which have different dependences on fundamental constants. Since both clocks are in the same general relativistic (GR) frame, any LPI violations that affect the two clocks differently can be unambiguously measured in the ratio. This may be parametrized, for example, by
\[
	\frac{\Delta (f_a/f_b)}{(f_a/f_b)} =(\beta_a - \beta_b)\frac{\Delta U}{c^2}
\]
where the parameters $\beta$ represent the response of each type of clock ($a$ and $b$) to a variation in gravitational potential $U$, which in turn is presumed to affect the fundamental constants (discussed in the next paragraph). In principle it may be possible to extract limits on variation of constants by comparison of two different clocks on two different satellites, however it cannot be done using only a single type of clock.

\paragraph{3.---}
After obtaining limits on the variation of the frequency ratio of two clocks, some theoretical calculation is required to interpret the result in terms of LPI violation. For example, to obtain a limit on the variation of $\alpha$ with variations in gravitational potential requires calculation of each clock's sensitivity to $\alpha$-variation. Earth-based LPI limits from comparison of cryogenic sapphire oscillator (CSO) clocks with hydrogen maser clocks~\cite{tobar10prd} find
\[
\beta_\textrm{H-maser} - \beta_\textrm{CSO} = -2.7\,(1.4) \times 10^{-4}
\]
from annual variations of the Earth's position in the Sun's gravitational field. This may be interpreted as a limit on the coupling of dimensionless fundamental constants to $U$. We define the coupling of $\alpha$ to $U$ by Eq.~(1) and similar equations for the dimensionless electron and light-quark mass ratios, $m_e/\Lambda_\textrm{QCD}$ and $m_q/\Lambda_\textrm{QCD}$, respectively, where $\Lambda_\textrm{QCD}$ is the quantum chromodynamics scale. Then $\beta_\textrm{H-maser} - \beta_\textrm{CSO} = 3 \kappa_\alpha + \kappa_e - 0.1\kappa_q$~\cite{flambaum04prd}, and we see that the terrestrial clock experiment places limits on this particular combination of coupling constants. Again, it is not meaningful or correct to say that cryogenic optical resonator clocks depend on the speed of light while atomic clocks depend on Planck's constant, as the experiment was interpreted by~\cite{kentosh12prl}.

We conclude that \cite{kentosh12prl} provides no new limits on variation of fundamental constants, and that improvements on existing limits using satellites will require satellites with more than one type of clock (e.g. cesium and rubidium clocks). Some experimental proposals are already underway in this direction~(see, e.g.,~\cite{schiller09ea}).

\end{document}